\documentclass[10pt,conference]{IEEEtran}
\IEEEoverridecommandlockouts

\usepackage{booktabs} 
\usepackage{makecell}
\usepackage{lipsum}
\usepackage[utf8]{inputenc}
\usepackage{array}
\usepackage{wrapfig}
\usepackage{multirow}
\usepackage{tabularx}
\usepackage{pifont}
\usepackage{lineno}
\usepackage{balance}
\usepackage{xcolor,pifont}
\usepackage{longtable}

\usepackage{soul}

\usepackage{balance}
\usepackage[linesnumbered, ruled,vlined]{algorithm2e}
\RequirePackage{pdflscape}
\usepackage{subcaption}
\usepackage{flushend}
\usepackage{multicol}
\usepackage{placeins}
\usepackage{booktabs}
\usepackage{indentfirst}
\usepackage{fancybox}
\usepackage{amssymb}
\usepackage[most]{tcolorbox}
\usepackage{fontawesome}
\usepackage{mathtools}
\usepackage{MnSymbol,wasysym}
\usepackage{fontawesome}
\usepackage{rotating}
\usepackage{adjustbox}
\usepackage{colortbl}
\usepackage{color}
 \usepackage{url}
\usepackage{framed}
\definecolor{Gray}{gray}{0.9}
\definecolor{shadecolor}{gray}{0.95}
\usepackage{tikz}
\usetikzlibrary{trees,positioning,shapes,shadows,arrows}
\usepackage[numbers,sort&compress]{natbib}

\newcommand{\findingboxx}[1]{
\begin{center}
\begin{tcolorbox}[colback=gray!11,
                  colframe=black,
                  boxrule=0.2mm,
                  width=0.45\textwidth,
                  arc=.8mm, auto outer arc,
                 ]
 #1
\end{tcolorbox}
\end{center}}

\tikzset{
  basic/.style  = {draw, text width=2cm, drop shadow, font=\sffamily, rectangle},
  root/.style   = {basic, rounded corners=2pt, thin, align=center, fill=white},
  level-2/.style = {basic, rounded corners=6pt, thin,align=center, fill=white, text width=3cm},
  level-3/.style = {basic, thin, align=center, fill=white, text width=1.8cm}
}
\usepackage{tcolorbox}

\newcommand{\todo}[1]{}

\begin{document}

\title{Computing Floating-Point Errors by Injecting Perturbations}
\author{
    Youshuai Tan\IEEEauthorrefmark{1}\IEEEauthorrefmark{4},
    Zhanwei Zhang\IEEEauthorrefmark{1}\IEEEauthorrefmark{4},
    Jinfu Chen\IEEEauthorrefmark{1}\IEEEauthorrefmark{5},
    Zishuo Ding\IEEEauthorrefmark{2},
    Jifeng Xuan\IEEEauthorrefmark{1}\IEEEauthorrefmark{5},
    Weiyi Shang\IEEEauthorrefmark{3}
    \thanks{\IEEEauthorrefmark{1}Wuhan University, China. Emails: tanyoushuai123@gmail.com, itbill8888@gmail.com, jinfuchen@whu.edu.cn, jxuan@whu.edu.cn}
    \thanks{\IEEEauthorrefmark{2}The Hong Kong University of Science and Technology (Guangzhou), China. Email: zishuoding@hkust-gz.edu.cn}
    \thanks{\IEEEauthorrefmark{3}University of Waterloo, Canada. Email: wshang@uwaterloo.ca}
    \thanks{\IEEEauthorrefmark{4}Both authors contributed equally to this research.}
    \thanks{\IEEEauthorrefmark{5}Corresponding authors.}
    
}


\maketitle

\begin{abstract}
Floating-point programs form the foundation of modern science and engineering, providing the essential computational framework for a wide range of applications, such as safety-critical systems, aerospace engineering, and financial analysis. Floating-point errors can lead to severe consequences. Although floating-point errors widely exist, only a subset of inputs may trigger significant errors in floating-point programs. Therefore, it is crucial to determine whether a given input could produce such errors.  Researchers tend to take the results of high-precision floating-point programs as oracles for detecting floating-point errors, which introduces two main limitations: (1) difficulty of implementation and (2) prolonged execution time. The two recent tools, ATOMU and FPCC, can partially address these issues. However, ATOMU suffers from false positives; while FPCC, though eliminating false positives, operates at a considerably slower speed.

To address these two challenges, we propose a novel approach named PI-detector to computing floating-point errors effectively and efficiently. Our approach is based on the observation that floating-point errors stem from large condition numbers in atomic operations (such as addition and subtraction), which then propagate and accumulate. PI-detector injects small perturbations into the operands of individual atomic operations within the program and compares the outcomes of the original program with the perturbed version to compute floating-point errors. We evaluate PI-detector with datasets from ATOMU and HSED, as well as a complex linear system-solving program. Experimental results demonstrate that PI-detector can perform efficient and accurate floating-point error computation.  

\end{abstract}


\section{Introduction}
\label{Introduction}

Floating-point programs are crucial to the advancement of science and engineering, powering applications in fields such as aerospace, finance, and data analysis~\cite{he2020testing, xu1994numerical, kotrla1996numerical, orszag1974numerical, rives1992joint, kobayashi1993modeling}. However, floating-point errors---the discrepancies between the exact mathematical value and the computed approximation of that value due to limitations in a computational process---are inevitable. Such errors in safety-critical systems like rocket launches and market fluctuations in financial applications can have profound impacts~\cite{zhang2024hierarchical}.

Previous researchers have developed various tools aimed at detecting inaccuracies in floating-point computations~\cite{benz2012dynamic, chowdhary2021parallel, chowdhary2022fast} to mitigate the potential impact of floating-point errors. A typical approach in floating-point error detection relies on the results of high-precision floating-point programs as oracles to verify the outputs of lower-precision floating-point programs. By comparing outputs from standard-precision computations against those from high-precision versions using identical inputs, tools can identify the errors~\cite{chiang2014efficient, zou2015genetic, yi2019efficient, guo2020efficient, wang2022detecting, zhang2023eiffel, zhang2024hierarchical}. While this method can be effective, it also presents several practical limitations: \textbf{1) difficulty of implementation:} High-precision floating-point program implementation often requires tailored operations that are specifically designed for various levels of precision, rather than a straightforward substitution of low-precision values with high-precision ones~\cite{wang2016detecting, zou2019detecting} (cf. Section~\ref{bac}). \textbf{2) prolonged execution time:} The computational overhead associated with high-precision programs can lead to substantially longer execution times, limiting their practicality in time-sensitive applications~\cite{larsson2013exploring}.

Currently, two advanced tools, ATOMU~\cite{zou2019detecting} and FPCC~\cite{yi2024fpcc}, have been developed to compute floating-point errors without the need for high-precision floating-point programs. ATOMU identifies floating-point errors by evaluating whether a given input can lead to large atomic condition numbers. However, ATOMU is prone to producing false positives~\cite{zou2019detecting} (cf. one example shown in Section~\ref{method}). FPCC addresses the false positives by calculating the condition number of the entire program, referred to as chain conditions. However, constructing chain conditions is extremely time-consuming (cf. Section~\ref{eva}), restricting the practical adoption of the approach.

To overcome such limitations, we propose a novel approach, named PI-detector (\textbf{P}erturbation \textbf{I}njection detector), to computing floating-point errors. PI-detector is grounded in the theory that floating-point errors stem from large atomic condition numbers~\cite{zou2019detecting, yi2024fpcc}. By injecting small perturbations into atomic operations, PI-detector can effectively expose latent bugs. To address limitation 1 (difficulty of implementation), we design PI-Detector as an LLVM pass. Our design significantly improves the ease of the detection process: users need only convert their original programs to LLVM IR and apply our pass during compilation. To evaluate whether PI-detector can successfully detect significant errors and overcome the limitation 2 (prolonged execution time), we apply PI-detector to the datasets from ATOMU~\cite{zou2019detecting} and HSED~\cite{zhang2024hierarchical}. Many numerical methods used to solve problems in engineering and science rely heavily on linear algebra, with the solution of general linear systems constituting a core aspect of this field~\cite{ford2014numerical}. Since existing datasets consist of simple functions, we take one classical linear system-solving program as a complex evaluation case to further assess PI-detector.

Our experimental results demonstrate that PI-detector can detect significant errors effectively and efficiently: \textbf{1) ATOMU and HSED datasets:} PI-detector produces only one false negative and 13 false positives, with all cases narrowly missing the significance threshold. The results indicate that PI-detector reliably identifies those significant errors. In comparison, FPCC generates fewer false positives (four) but significantly more false negatives (34). FPCC and the high-precision approach need about \(10\times\) and \(408\times\) the execution time of PI-detector. \textbf{2) Complex floating-point programs: } The errors calculated by PI-detector show a strong correlation with those computed by the high-precision approach, while requiring only 1/45 of the runtime. In contrast, FPCC cannot handle programs with multiple inputs, making it unsuitable for deployment on our complex datasets. This demonstrates PI-detector’s strong practical applicability. The replication package, including the datasets and scripts, is available on the link\footnote{https://github.com/anony-repository/PI-detector}.

In summary, our work makes the following three main contributions:
\begin{itemize}
\item We introduce PI-detector, a novel approach designed to efficiently compute floating-point errors in floating-point programs.
\item We evaluate PI-detector on a diverse set of benchmarks, including both popular, i.e., ATOMU and HSED, and complex floating-point programs.
\item We conduct experiments to demonstrate that our approach performs better than high-precision approach, ATOMU, and FPCC.
\end{itemize}

To clarify key concepts used throughout this manuscript, we provide the following definitions: In this context, we define the \textbf{high-precision approach} as the approach that uses the results of high-precision programs as oracles to calculate floating-point errors. We refer to \textbf{high-precision programs} as high-precision floating-point programs. Following ATOMU, we use the term \textbf{significant errors} to denote floating-point bugs whose relative errors exceed 0.001.

\textbf{Paper organization.} The remainder of the paper is organized as follows. Section~\ref{bac} presents the background of our work. We introduce the limitations of the existing approaches in Section~\ref{motivate}. Section~\ref{method} elaborates on PI-detector. The evaluation of our approach is presented in Section~\ref{eva}. We discuss limitations and introduce related work in Section~\ref{threat} and Section~\ref{related-work}, respectively. Finally, Section~\ref{conclusion} concludes our work.

\section{Background}
\label{bac}
In this section, we provide an overview of the background knowledge related to floating-point error measurement, condition number, and matrix condition number. 

\subsection{Floating-point error measurement}
Floating-point numbers are the standard numerical representation used in modern computers to simulate real numbers. However, due to the inherent approximation and errors of floating-point arithmetic~\cite{jezequel2008cadna}, results produced by floating-point programs are prone to errors compared to the outcome of mathematical operations over the field of real numbers. For instance, in Java SE Development Kit 21.0.3, subtracting the floating-point number 1.1 from 1.2 does not yield 0.1; instead, it results in 0.09999999999999987. In this context, we denote the real number as \(x\) and its floating-point representation as \(\hat{x}\). With respect to a floating-point program \(\textbf{P: }\hat{y}=\hat{f}(\hat{x})\), \(\hat{y}\) refers to the floating-point result and \(f(x)\) represents for the real outcome. The error  between the ideal mathematical result \(f(x)\) and the outcome of the floating point program \(\hat{f}(\hat{x})\) can be quantified using the following two common metrics:

\begin{equation}
\begin{aligned}
& Err_{abs}(f(x),\hat{f}(\hat{x}))=\left|f(x)-\hat{f}(\hat{x})\right|\quad \\
& Err_{rel}(f(x),\hat{f}(\hat{x}))=\left|\frac{f(x)-\hat{f}(\hat{x})}{f(x)}\right|
\end{aligned}
\end{equation}
where \(Err_{abs}\) and \( Err_{rel}\) stand for absolute error and relative error, respectively. However, if the value of \(f(x)\) approaches 0, it can lead to a division-by-zero error. 

To mitigate this issue, Unit in the Last Place \((ULP)\) is often used~\cite{zou2019detecting}. 
\(ULP(x)\) is the gap between a given floating-point number \(x\) and the next representable floating-point number~\cite{overton2001numerical}. The formulation for \(ULP\) is given by:

\begin{equation}
ULP(x)=\epsilon\times2^E
\end{equation}
where \(\epsilon\) is the machine epsilon and \(E\) represents for the exponent of \(\hat{x}\). Therefore, different numbers possess varying magnitudes of \(ULP\) and \( Err_{ulp}\) is flexible:

\begin{equation}
Err_{ulp}(f(x),\hat{f}(\hat{x}))=\left|\frac{f(x)-\hat{f}(\hat{x})}{ULP(f(x))}\right|
\end{equation}

\subsection{Condition number}
Condition number serves as a crucial indicator of how the difference in the input \(x\) is amplified during 
the operation of \(f(x)\)~\cite{overton2001numerical}. Its derivation can be elucidated using the Taylor Expansion Theorem~\cite{zou2019detecting}. The difference in \(f(x)\) at a perturbed value \(x + \Delta x\) can be expressed as:

\begin{equation}\begin{aligned}
&=\left|\frac{f(x+\Delta x)-f(x)}{f(x)}\right| \\
&=\left|\frac{f(x+\Delta x)-f(x)}{\Delta x}\cdot\frac{\Delta x}{f(x)}\right| \\
&=\left|(f^{\prime}(x)+\frac{f^{\prime\prime}(x+\theta\Delta x)}{2!}\Delta x)\cdot\frac{\Delta x}{f(x)}\right|,\theta\in(0,1) \\
&=\left|\frac{\Delta x}{x}\right|\cdot\left|\frac{xf^{\prime}(x)}{f(x)}\right|+O\big((\Delta x)^{2}\big) \\
&=Err_{rel}(x,x+\Delta x)\cdot\left|\frac{xf^{\prime}(x)}{f(x)}\right|+O\big((\Delta x)^2\big)
\end{aligned}
\end{equation}
where \(\Delta x\) represents for small perturbation and \(\theta\) is the Lagrange form of the remainder. Consequently, the equation derives the formula for the condition number \cite{higham2002accuracy}, \(\left|\frac{xf^{\prime}(x)}{f(x)}\right|\). Since we can ignore the term \(O\big((\Delta x)^2\big)\), we can conclude that a minor difference of the input could result in a significant difference in the output of programs if the condition number is large. The conversion of input values into floating-point representations introduces inherent errors, which means that the inputs to floating-point programs inherently contain inaccuracies. Thus, by determining the condition number of a given program for a specific input, we can assess whether the corresponding output is likely to exhibit large errors. However, condition number contains the derivative of the function \(f\) with respect to \(x\). As a result, calculating the condition number can be time-consuming for complex programs~\cite{fu2015automated}. 

As programs consist of atomic operations, ATOMU~\cite{zou2019detecting} lists common atomic operations and provides their corresponding condition numbers. Table~\ref{condition} lists the cases which could lead to significant condition numbers.

\begin{table}[tb]
\centering
\footnotesize
\caption{Condition numbers and dangerous regions that are likely to produce large errors~\cite{zou2019detecting}.}
\label{condition}
\resizebox{0.49\textwidth}{!}{
\begin{tabular}{lcr}
\toprule
 \textbf{Operation (\(op\))} & \textbf{Condition Number (\(C_{op}\))} & \textbf{Dangerous Region}\\
\midrule
\(op(x,y)=x+y\)   &  \(C_{+,x}(x,y)=\left|\frac{x}{x+y}\right|, C_{+,y}(x,y)=\left|\frac{y}{x+y}\right|\) & \(x\approx-y\)\\
  \(op(x,y)=x-y\)   & \(C_{-,x}(x,y)=\left|\frac{x}{x-y}\right|,C_{-,y}(x,y)=\left|-\frac{y}{x-y}\right|\)  & \(x\approx y\) \\
\(op(x)=\sin(x)\) &  \(C_{\sin}(x)=|x\cdot\cot(x)|\) & \(x\to n\pi,n\in\mathbb{Z} \setminus \{0\}\) \\
\(op(x)=\cos(x)\) & \(C_{\cos}(x)=|x\cdot\tan(x)|\) & \(x\to n\pi+\frac\pi2,n\in\mathbb{Z}\)   \\
\(op(x)=\tan(x)\) & \(C_{\tan}(x)=\left|\frac x{\sin(x)\cos(x)}\right|\) & \(x\to\frac{n\pi}2,n\in\mathbb{Z} \setminus \{0\}\) \\
\(op(x)=\arcsin(x)\) & \(C_{\arcsin}(x)=\left|\frac{x}{\sqrt{1-x^2}\cdot\arcsin(x)}\right|\) & \(x\to-1^+,x\to1^-\) \\
\(op(x)=\arccos(x)\) & \(C_{\arccos}(x)=\left|-\frac{x}{\sqrt{1-x^2}\cdot\arccos(x)}\right|\) & \(x\to-1^+,x\to1^-\) \\
\(op(x)=\sinh(x)\)  & \(C_{\sinh}(x)=|x\cdot\coth(x)|\) & \(x\to\pm\infty \) \\
\(op(x)=\cosh(x)\) & \(C_{\cosh}(x)=|x\cdot\tanh(x)|\) &   \(x\to\pm\infty \) \\
\(op(x)=\exp(x)\) & \(C_{\exp}(x)=|x|\) &   \(x\to\pm\infty \)\\
\(op(x)=\log(x)\) & \(C_{\log}(x)=\left|\frac1{\log x}\right|\) & \(x \to 1\) \\
\(op(x)=\log_{10}(x)\) & \(C_{\log10}(x)=\left|\frac1{\log x}\right|\) & \(x \to 1\) \\
\(op(x,y)=x^y\)  & \(C_{pow,x}(x,y)=|y|,C_{pow,y}(x,y)=|y\log(x)|\) & \(x\to0^+,y\to\pm\infty \) \\

\bottomrule
\end{tabular}
}
 \vspace{-0.5cm}
\end{table}

\subsection{Matrix condition number}
In the context of numerical linear algebra, the matrix condition number is a fundamental concept~\cite{ford2014numerical}. It quantifies the sensitivity of the solution of a linear system of equations to small changes in the coefficient matrix or the right-hand side vector. The number \(\left\|A\right\|\left\|A^{-1}\right\|\) is the matrix condition number of the matrix \(A\), denoted by \(\kappa \left(A\right)\). If \(\kappa \left(A\right)\) is large, achieving an accurate solution for a linear system becomes increasing difficult~\cite{ford2014numerical} because the inherent floating-point errors of inputs are amplified. For instance, considering the matrix \(A=\begin{bmatrix}1.0001&1\\1&1\end{bmatrix}\), this matrix has a large condition number. The large condition number indicates that the linear system \(Ax=b\) is ill-conditioned. A small perturbation in the right-hand side vector \(b\) can lead to a significant change in the solution vector \(x\).
Specifically, the solution of the linear systems \(Ax\)=\(\left[{\begin{array}{cc}1.0001&1\\1&1\end{array}}\right]\left[{\begin{array}{c}x_{1}\\x_{2}\end{array}}\right]=\left[{\begin{array}{cc}2.0001\\2\end{array}}\right]\) and \(Ax\)=\(\left[\begin{array}{cc}1.0001&1\\1&1\end{array}\right]\left[\begin{array}{c}x_{1}\\x_{2}\end{array}\right]=\left[\begin{array}{c}2\\2\end{array}\right]\) yield outputs that differ significantly, despite the small perturbation in \(b\). In the first case, the results are \(x_1 = 1\) and \(x_2 = 1\), whereas in the second case, they are \(x_1=0.0000\) and \(x_2=2.0000\). This example illustrates how even minimal changes in input can lead to substantial variations in the output when dealing with matrices that possess a high condition number. 

The observed inaccuracies are attributed to floating-point errors within the programs. Since existing datasets primarily consist of simple functions, we employ linear system-solving programs as representative complex programs to further evaluate the effectiveness of PI-detector.

\section{Motivation}
\label{motivate}

Due to the potentially severe consequences of floating-point errors, many techniques~\cite{chiang2014efficient, zou2015genetic, zou2019detecting, yi2019efficient, guo2020efficient, wang2022detecting, zhang2023eiffel, zhang2024hierarchical} have been proposed to identify inputs that could lead to significant errors in floating-point programs. These techniques rely on search-based algorithms, which necessitate oracles for guidance. Most of the existing tools utilize high-precision programs to obtain oracles. However, the use of high-precision programs contains two inherent limitations.

\begin{figure}[htbp]
\vspace{-0.4cm}
    \centering
    \includegraphics[width=0.48\textwidth]{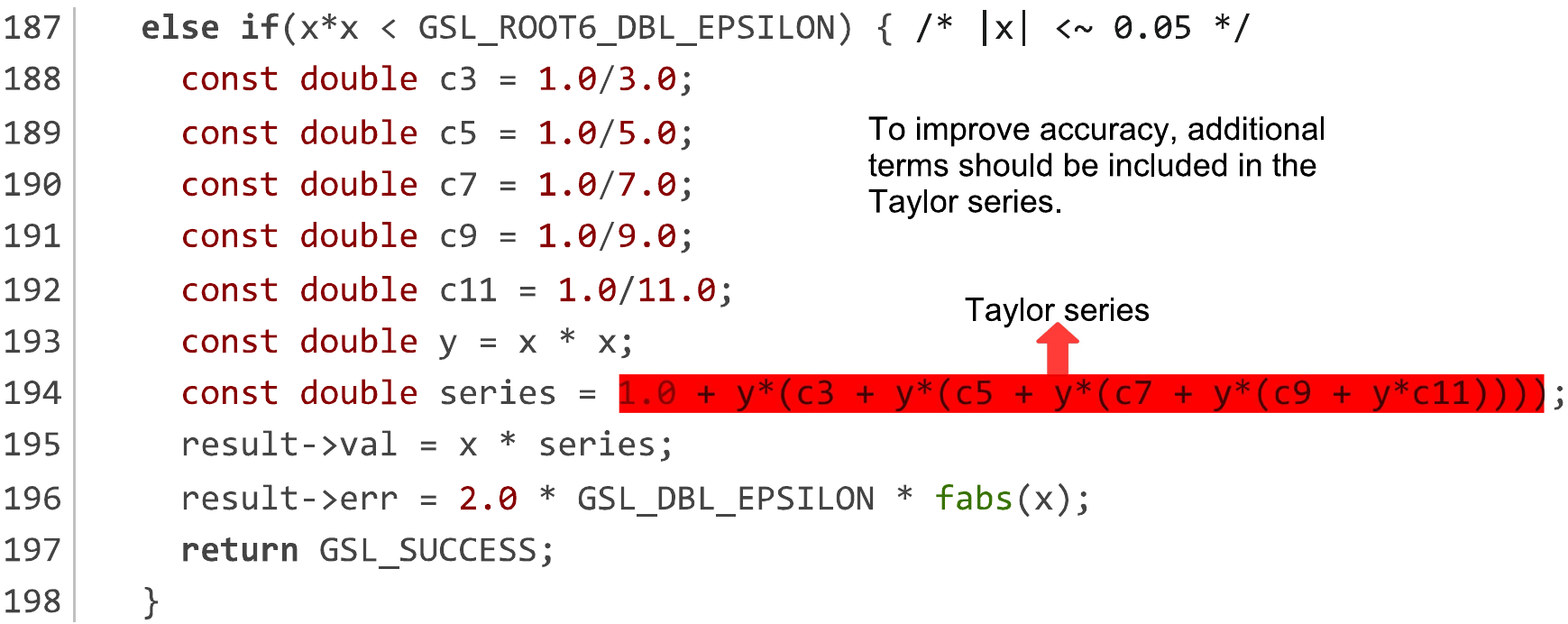}
    \caption{Code snippet of \textit{gsl\_sf\_legendre\_Q0\_e}.}
    \vspace{-0.3cm}
    \label{legendre}
\end{figure}

\textbf{Limitation 1: Difficulty of implementation.} Developing high-precision programs requires specialized knowledge of mathematics and numerical analysis. High-precision programs often involve precision-specific operations rather than merely substituting low-precision numbers with high-precision ones~\cite{wang2016detecting, zou2019detecting}. For example, Figure~\ref{legendre} illustrates one branch of the GSL function\footnote{~\url{https://www.gnu.org/software/gsl/doc/html/}} \textit{gsl\_sf\_legendre\_Q0\_e}, which is implemented based on a Taylor series expansion. Simply increasing the precision of the floating-point representation does not lead to significantly more accurate results, due to the truncation of higher-order terms in the Taylor series. Therefore, achieving higher accuracy requires not only improving the numerical precision but also including more terms in the Taylor series.

\textbf{Limitation 2: Prolonged execution time.} In terms of runtime performance, quadruple precision (128 bits) is about 100 times slower than double precision (64 bits)~\cite{larsson2013exploring}. Additionally, programs that use arbitrary precision libraries, such as MPFR~\cite{fousse2007mpfr} and \textit{mpmath}~\cite{mpmath}, face even greater slowdowns as they increase precision levels.

 We believe that the two limitations can have significant negative impacts on software tasks, such as floating-point software testing. Firstly, the complexity of rewriting high-precision programs to obtain oracles significantly raises the barrier to testing. Secondly, the lengthy execution time of high-precision programs renders them impractical for testing complex floating-point software. 
 
Recently, two methods---ATOMU and FPCC---have been proposed to identify inputs that lead to significant errors. Unlike traditional approaches, they do not rely on high-precision programs to determine whether a given input can cause substantial error. However, ATOMU detects significant errors by directly assessing whether the input can induce large condition numbers across all operations in the program, which may lead to false positives. To mitigate this issue, FPCC computes the condition number of the entire program and uses it as a more reliable indicator for error detection. Nevertheless, this method incurs excessively high computational overhead.

 To address the limitations, we propose PI-detector, a simple yet effective approach for computing floating-point errors.
\section{PI-detector: computing errors by injecting perturbations into floating-point programs}
\label{method}

PI-detector is designed to provide the software engineering community with an accurate and efficient tool for computing floating-point errors. The arguments of PI-detector are a floating-point program and its corresponding inputs. First, we generate LLVM Intermediate Representation (IR)~\cite{lattner2004llvm} of the target program. Next, we generate perturbed LLVM IR by injecting small perturbations into the original LLVM IR. Finally, we use two LLVM IRs and the input to compute two results, and the difference between the two results is considered as the error. 

\begin{figure*}[htbp]
\centerline{\includegraphics[width=0.95\textwidth]{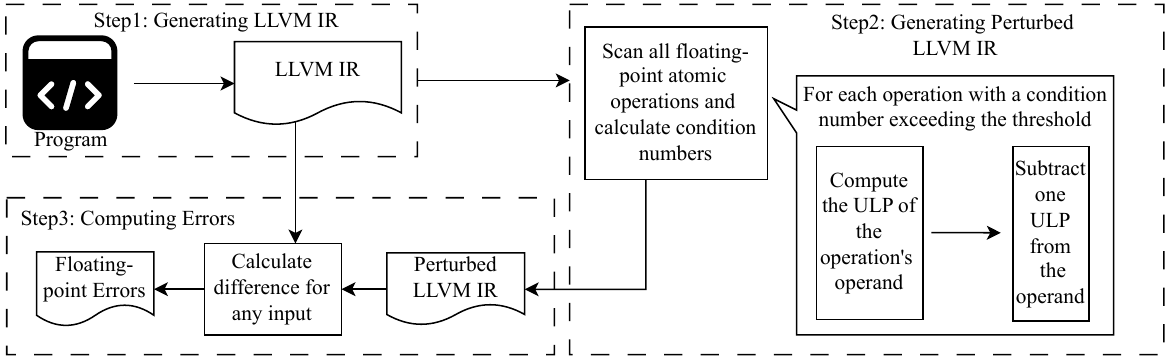}}
\caption{An overview of PI-detector.}
\vspace{-0.3cm}
\label{framework}
\end{figure*}

In this section, we first introduce the theoretical basis and design inspiration of PI-detector. Subsequently, we elaborate on PI-detector, which comprises three main steps. We present a floating-point program as a representative example to explain the origin of floating-point errors and illustrate our approach. An overview of our approach is shown in Figure~\ref{framework}.

\subsection{Foundations of PI-detector}
Before presenting the details of our approach, PI-detector, in this subsection, we first introduce the theoretical basis underlying our approach---namely, that significant floating-point errors arise due to large condition numbers. Subsequently, we describe the origin of the idea behind computing errors through error injection.

\noindent \textbf{Theoretical basis: Large condition numbers of atomic operations magnify the inherent floating-point errors in the operands.} The condition number is given by the relation \(Err_{rel}(f(x),f(x+\Delta x))=Err_{rel}(x,x+\Delta x)\cdot\left|\frac{xf^{\prime}(x)}{f(x)}\right|\) (cf. Section~\ref{bac}). This formula illustrates that computational results can incur substantial errors when the operands themselves contain errors and the condition number is large. In floating-point arithmetic, errors are introduced at the very outset when inputs (real numbers) are approximated in finite precision. Subsequently, every arithmetic of the program will also produce error due to the floating-point arithmetic characteristic. Consequently, every variable in a floating-point program carries some degree of error. If any operation within the algorithm has a large condition number, these errors may be significantly amplified, leading to unreliable results. As demonstrated in Table~\ref{method_}, the subtraction operation exhibits a significantly large condition number, leading to a substantial floating-point error.

\noindent \textbf{Design inspiration: How to trigger the large condition number and avoid false positives simultaneously.} Based on the aforementioned condition number theory, ATOMU detects errors by assessing whether a large condition number exists. However, naively relying on this criterion alone can lead to false positives. For the last operation of our example in Table~\ref{method_}, the error is negligible although the operand contains a large error. To address this challenge, our design philosophy centers on two key objectives: 1) reliably triggering large condition numbers to expose potential error amplification, and 2) dynamically tracking error propagation to eliminate false positives. We achieve this through a systematic perturbation-based approach. By deliberately injecting perturbations into the operands of each operation whose condition number is large, we establish the following mechanisms: \textbf{1) Error Amplification Detection:} If an operation has a large condition number, the perturbation will induce a measurable discrepancy between the original and perturbed computational paths. \textbf{2) Error Cancellation Identification:} If all variables with large errors disappear (such as the third operation of our example in Table~\ref{method_}), the final result of the perturbed program will not be substantially changed by these variables, thereby avoiding false positives.

\begin{table*}[tb]
\centering
\footnotesize
\caption{An example illustrating the error calculation of PI-detector and the input argument is 1.3694384060045659. The example contains three atomic operations. }
\label{method_}

\resizebox{\linewidth}{!}{
\begin{tabular}{lccccr}
\toprule
\textbf{Operation}  & \textbf{Original} & \textbf{Perturbed} & \textbf{Error from} & \textbf{Error from} & \textbf{Condition}\\
\textbf{}  & \textbf{Program} & \textbf{Program} & \textbf{PI-detector} & \textbf{high-precision program} & \textbf{Number}\\
\midrule
 \(\cos(x)\) & 1.9999999999999993e-01 & 1.9999999999999993e-01 & 0 & 5.0437e-02  & 6.7089e+00\\
 \([\cos(x)] - 0.2\) &     -8.326672684688674e-17 &  -1.1102230246251565e-16    &  2.2518e+15  & 1.0143e+15   & 2.4019e+15\\
\( [\cos(x) - 0.2] + 10 \) & 10 &  10  & 0 & 3.9837e-02  & 8.3267e-18\\

\bottomrule
\end{tabular}
}
\end{table*}

\subsection{Workflow of PI-detector}
In this subsection, we present the details of systematic perturbation-based approach. To ensure both usability and computational efficiency, we implement the perturbation operations as an LLVM pass. We present the workflow of our approach and demonstrate how PI-detector avoids false positives using our illustrative example.

\begin{figure}[htbp]
 \vspace{-0.5cm}
    \centering
    \begin{minipage}{0.49\textwidth} 
        \centering
        \includegraphics[width=\textwidth]{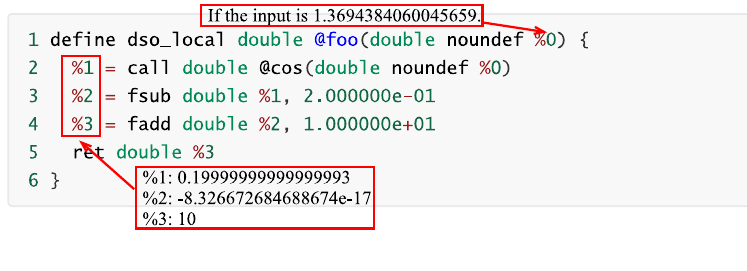}
        \vspace{-0.9cm}
        \subcaption{Original IR}\label{original_ir}
    \end{minipage}
    
    \begin{minipage}{0.49\textwidth} 
    
        \centering
        \includegraphics[width=\textwidth]{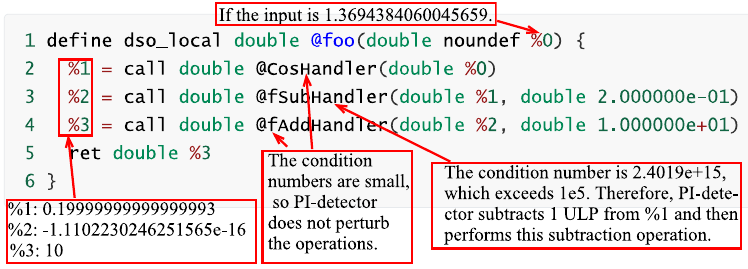}
        \subcaption{Perturbed IR}\label{perturbed_ir}
    \end{minipage}
    \caption{LLVM IR of the illustrative example.} 
    \label{ir}
\vspace{-0.8cm}
\end{figure}

\noindent \textbf{Step 1: Generating LLVM IR of the original programs.} This initial step involves compiling floating-point programs into LLVM IR, enabling us to leverage LLVM's optimizer to introduce perturbations directly into the program, rather than requiring specialized analysis for each individual case. Moreover, many programming languages, such as C, C++, Go, and Python, could generate LLVM IR based on the source code~\cite{sarda2015llvm}. As a result, PI-detector has the potential to serve as a general-purpose tool applicable to a wide range of programs. Figure~\ref{original_ir} shows the LLVM IR of our illustrative example.

\noindent \textbf{Step 2: Generating perturbed LLVM IR by subtracting one ULP from the operands of each operation with a large condition number.} The perturbation injection is implemented as an LLVM pass in PI-detector. First, PI-detector scans the LLVM IR instruction by instruction. Afterwards, our approach injects a function call whenever it encounters floating-point atomic operations. Next, the function calculates the condition numbers of all the floating-point operations. If the condition number exceeds a predefined threshold (set to 1e5 following ATOMU), PI-detector perturbs this operation. Specifically, PI-detector computes the ULP of the atomic operation's operand and returns the modified operand by subtracting one ULP. As the errors of elementary arithmetic operations are bounded by an upper limit~\cite{zou2019detecting}, i.e., 1 + 1/1000 ULPs, we set the value of perturbation to one ULP. Note that the perturbation is applied exclusively to the operand that results in large condition numbers. Figure~\ref{perturbed_ir} illustrates the perturbed LLVM IR for the illustrative example.

\noindent \textbf{Step 3 (addressing limitation 1 and limitation 2): Computing floating-point errors by calculating the difference between the results of the original program and the perturbed version.} After injection, the instrumented LLVM IR is compiled into a library. Any client program can use this library to obtain the error by calculating the difference between the original program and the perturbed one. On the contrary, developers have to spend much effort and time designing high-precision programs when applying the high-precision approach. Therefore, PI-detector can solve the limitation 1 (difficulty of implementation). PI-detector incurs minimal overhead compared to the high-precision approach because it only performs three additional operations, i.e., condition number computation, ULP calculation, and the modified result computation. Therefore, this efficient approach ensures that PI-detector remains a practical solution without excessive computational cost, and addresses the limitation 2 (prolonged execution time).

\noindent \textbf{An illustrative example that demonstrates our approach.} We use the example to demonstrate PI-detector's workflow and show how it avoids the false positives of ATOMU. Figure~\ref{ir} records the intermediate results of the function's execution taking 1.3694384060045659 as the input. In the second step, the condition number of \(\%1 - 0.2\) respect to \(\%1\) is \(\left|\frac{0.19999999999999993}{0.19999999999999993-0.2}\right|\) whose value is 2.4019e+15. As discussed in Section~\ref{bac}, big condition number can result in significant floating-point errors by amplifying operands' errors. When using the result from a high-precision program as an oracle, the ULP error of \%2 is about 1.0143e+15, which aligns with this theory. The ULP difference between \%2 of original program and the perturbed version is 2.2518e+15, demonstrating that PI-detector can find the error by injecting perturbations. In the final step, the ULP error of the final result is zero because of adding a small number to a large number. As a result, ATOMU would incorrectly classify this input as positive. In contrast, PI-detector can avoid this mistake by executing all the operations. The ULP difference between \%3 of the original program and the perturbed one is zero, highlighting PI-detector's ability to accurately identify errors.

\section{Evaluation}
\label{eva}
In Section~\ref{method}, by treating target floating-point programs as black boxes, we have demonstrated that our approach does not require a complex implementation to address \textbf{limitation 1 (difficulty of implementation)}. In this section, we perform empirical evaluations to assess PI-detector’s effectiveness and efficiency using multiple datasets. 

We begin our evaluation by presenting the experimental setup, including the datasets, evaluation metric, and our experimental environment. Subsequently, we report the experimental results to demonstrate that PI-detector can solve the two limitations of high-precision approach: difficulty of implementation and prolonged execution time. Since FPCC addresses the false-positive issue of ATOMU and is the state-of-the-art approach that does not rely on high-precision programs, we adopt it as the baseline approach. Our results also show that our approach not only reduces false positive cases in ATOMU but also achieves higher efficiency compared to FPCC.

\subsection{Experiment setup}

\subsubsection{Dataset}

We base our evaluation on datasets from the open-source packages of ATOMU~\cite{zou2019detecting} and HSED~\cite{zhang2024hierarchical}, both designed to find inputs that lead to significant errors. First, we need a set of verified significant errors to assess whether our approach can successfully detect them. ATOMU considers a relative error greater than 0.1\% as significant and obtains some inputs. Therefore, we take these 172 cases as our evaluation data. For HSED, we also select the cases whose relative errors are more than 0.1\%, resulting in two cases. In total, we use 174 cases with verified significant errors for evaluation, while the remaining 899 cases are non-significant.

Existing research tends to focus on small floating-point programs, such as GSL functions. We believe that the lack of datasets for complex functions is due to the difficulty of identifying inputs that lead to significant errors in such functions. This is largely because current techniques for finding such inputs have several limitations, including their inability to handle multi-input functions, inefficient error computation, and other shortcomings. To evaluate the potential applicability of PI-detector for more complex programs, we take the linear system-solving program as a complex test case. If the condition number of the coefficient matrix of a linear system is large, accurately solving the system becomes significantly challenging~\cite{ford2014numerical} (cf. Section~\ref{bac}). Furthermore, the condition number of a singular matrix is infinite. Therefore, by making the coefficient matrix of linear systems approach singularity, we can construct linear system-solving programs that produce large floating-point errors. This allows us to easily generate datasets containing complex, long-running programs that produce large floating-point errors.

\subsubsection{Evaluation metric}
We utilize \(Err_{ulp}\) (c.f. Section~\ref{bac}) and \(Err_{rel}\) to quantify floating-point errors. Our approach takes \(Err_{abs}\) as the difference between the results of the original program and the perturbed one, i.e. \(\left|Res_{Ori}-Res_{Per}\right|\). \(Err_{rel}\) and \(Err_{ulp}\) are \(\left|Err_{abs} / Res_{Ori}\right|\) and \(\left|Err_{abs} /ULP(Res_{Ori})\right|\), respectively. FPCC only provides chain condition number, we multiply the relative error of the input by it to obtain the \(Err_{rel}\) according to the condition number formula (cf. Section~\ref{bac}). Since our approach takes one ULP as perturbation to trigger errors, we also set \(Err_{rel}(x,x+\Delta x)\) as \(\left|\frac{ULP(x)}{x}\right|\).

We use Mann-Kendall test~\cite{mann1945nonparametric} to detect monotonic trends. For correlation analysis, we calculate both the Pearson and Spearman correlation coefficients to assess the association between errors computed by PI-detector and high-precision approach. The Pearson correlation coefficient measures the strength of the linear association between two variables~\cite{sedgwick2012pearson}. The Spearman correlation coefficient, on the other hand, is defined as the Pearson correlation coefficient between the rank variables~\cite{myers2013research}, assessing monotonic relationships.

\subsubsection{Software and hardware environment}
Our primary experimental environment is a Docker container running Ubuntu 24.04 on a laptop equipped with an AMD Ryzen 7 6800H @ 3.20 GHz CPU and 16GB RAM. However, for solving systems of linear equations, we employ a separate Docker container with Ubuntu 22.04 on a system with a 13th Gen Intel(R) Core(TM) i9-13900K CPU and 128GB RAM to accommodate higher computational demands.

\subsection{Evaluation results}
In this subsection, we report the experimental results of evaluating our proposed approach, PI-detector, organized around two research questions (RQs). For each RQ, we first describe the underlying motivation and evaluation approach, followed by a detailed discussion of the results.

\subsection*{RQ1 (mitigating the issue of false positives from ATOMU): Are floating-point errors computed by PI-detector reliable?}
\label{rques1}

\begin{figure}[htbp]
    \centering
    \includegraphics[width=0.48\textwidth]{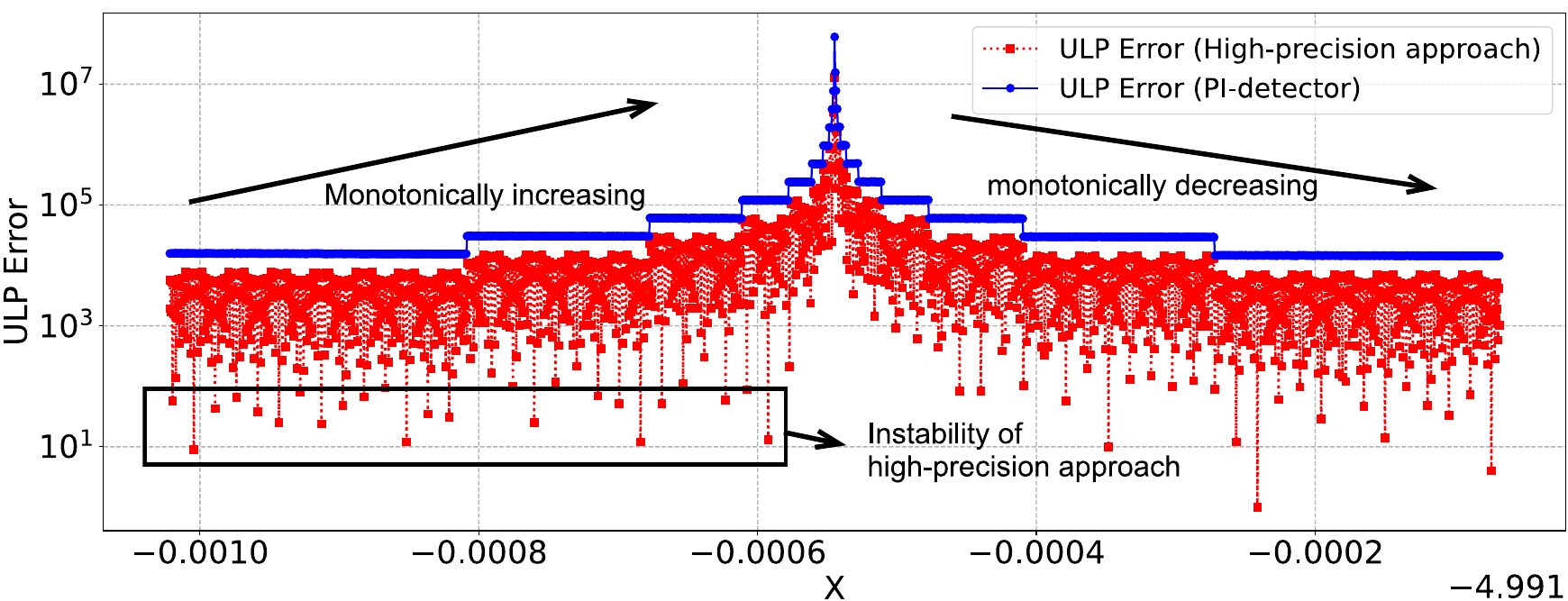}
     \vspace{-0.2cm}
    \caption{Comparison of errors computed by PI-detector and the high-precision approach in a small region containing significant error of \textit{gsl\_sf\_lngamma\_e}.}
    \vspace{-0.5cm}
    \label{rq1}
\end{figure}

\noindent\textit{\textbf{Motivation.}} Calculating the floating-point error of the given input and program is the basis of many topics of software engineering community, such as detecting the inputs that lead to significant errors for floating-point programs~\cite{chiang2014efficient, zou2015genetic, yi2019efficient, guo2020efficient, wang2022detecting, zhang2023eiffel, zhang2024hierarchical, zou2019detecting, yi2024fpcc}. To assess the effectiveness of PI-detector, we evaluate its capability to identify significant errors by measuring True Positives (TP), False Positives (FP), False Negatives (FN), and True Negatives (TN). As discussed in Section~\ref{bac}, the condition number grows as inputs approach the dangerous region, where floating-point errors become more pronounced. Therefore, the floating-point error increases when the input approaches the point with the local maximum floating-point error. Within a specific interval, the floating-point error should initially increase monotonically and then decrease monotonically, as shown in Figure~\ref{rq1}. Since the monotonicity is useful in finding the local maximum floating-point error~\cite{yi2024fpcc}, we further examine whether PI-detector correctly captures this trend in its error estimations.

\noindent\textit{\textbf{Approach.}} We conduct our evaluation on three datasets to verify whether PI-detector can effectively detect significant errors. \textbf{(1) ATOMU and HSED.} First, we compute the relative errors for all 174 significant-error cases across the two datasets using our approach to assess whether our approach can detect the 174 significant errors. We compute False Positives (FP), False Negatives (FN), and True Negatives (TN) based on the remaining 899 cases that result in non-significant errors. Second, we extend the set of points that produce significant errors to assess whether the errors computed by our approach exhibit the expected trend. Specifically, we extend 1,000 points to both the left and right for the cases of ATOMU, using intervals of ten times the 64-bit ULP. We exclude cases where the absolute value exceeds \(10^9\), as the corresponding ULPs are too large to reveal local trends. Additionally, for cases where values are particularly close to zero, we appropriately expand the range. The interval used for the two cases of HSED is one 32-bit ULP, as the programs are relatively simple and less sensitive to changes in input. As shown in Figure~\ref{rq1}, based on the underlying principle of error generation, the region around the point with the maximum error is expected to exhibit a monotonic pattern: the left side should be monotonically increasing, while the right side should be monotonically decreasing. Therefore, we divide the extended region into two halves and apply the Mann-Kendall test separately to assess monotonicity, requiring a p-value less than 0.05 for significance. Moreover, as illustrated in Figure~\ref{rq1}, the errors computed by the high-precision approach are not consistently stable—occasionally being very small—which results in deviations from strict monotonicity. Thus, to further quantify this monotonic behavior of errors computed by different approaches, we employ the Mann-Kendall statistic \(S\) as a metric~\cite{machida2013effectiveness}. A larger absolute value of \(S\) indicates a more pronounced monotonic trend. \textbf{(2) 3,000-dimensional linear system-solving program.} To ensure that the linear system-solving program's runtime is neither too long nor too short, we chose 3,000-dimensional matrices as the coefficient matrices for the experimental data. We generate a dataset of 1,000 matrices, some of which are randomly modified to be nearly singular to exhibit huge matrix condition numbers and numerical instability (i.e., large floating-point errors). We use the \textit{lu\_solve} function of GSL to implement normal-precision linear system solving. For the high-precision program, we employ the \textit{Rgesv} function of \textit{mplapack}\footnote{https://github.com/nakatamaho/mplapack} with 128-bit floating point numbers, which also performs the LU Decomposition to solve linear systems. Since the significant error threshold, i.e., 0.1\%, is designed for the simple functions, we can not utilize it for complex ones. Thus, we compute correlations between floating-point errors calculated by our approach and the high-precision approach to evaluate the effectiveness of PI-detector.

\noindent\textit{\textbf{Results. }}\textbf{PI-detector only misses one significant error from the ATOMU and HSED datasets.} Table~\ref{error_and_correlation} presents the relative errors obtained from the three approaches. Following ATOMU~\cite{zou2019detecting}, we set the significance threshold at 0.001 to classify errors. Since the high-precision approach serves as the ground truth, all its derived errors necessarily exceed 0.001. PI-detector only miss one significant error, but the error (\(7.47e-04\)) is within a negligible margin of the threshold. In contrast, FPCC fails to detect 34 significant errors. To further evaluate detection accuracy, we record false positives (FP) and true negatives (TN) of the two approaches. Our approach yields 13 FP and 855 TN, whereas FPCC produces 4 FP and 864 TN. It is noted that the average and minimum errors calculated by the high-precision approach for the PI-detector's 13 FP are 4.94E-04 and 2.38E-05, respectively. The values demonstrate that PI-detector's FP closely approaches the significance threshold (i.e., 0.001). Therefore, if the threshold is not strictly enforced, the results obtained by our approach are consistent with those produced by the high-precision approach. 

\begin{table*}[]
\caption{Errors calculated by our approach, high-precision approach, and FPCC. (Note: \textbf{Err.} = Error, \textbf{App.} = Approach, \textbf{Hig.} = High-precision) Values in bold denote those below 0.001.} 
\vspace{-0.2cm}
\scriptsize
\label{error_and_correlation}
\centering
\setlength{\tabcolsep}{5pt}
\begin{tabular}{lrrrlrrrlrrr}
\cmidrule(lr){1-4} \cmidrule(lr){5-8} \cmidrule(lr){9-12} 
\multicolumn{1}{c}{\multirow{2}{*}{\textbf{\begin{tabular}[c]{@{}c@{}}Function\\ Index\end{tabular}}}} & \multicolumn{1}{c}{\multirow{2}{*}{\textbf{\begin{tabular}[c]{@{}c@{}}Err. of\\ Our App.\end{tabular}}}} & \multicolumn{1}{c}{\multirow{2}{*}{\textbf{\begin{tabular}[c]{@{}c@{}}Err. of \\ Hig. App.\end{tabular}}}} & \multicolumn{1}{c}{\multirow{2}{*}{\textbf{\begin{tabular}[c]{@{}c@{}}Err. of\\ FPCC\end{tabular}}}} & \multicolumn{1}{c}{\multirow{2}{*}{\textbf{\begin{tabular}[c]{@{}c@{}}Function\\ Index\end{tabular}}}} & \multicolumn{1}{c}{\multirow{2}{*}{\textbf{\begin{tabular}[c]{@{}c@{}}Err. of\\ Our App.\end{tabular}}}} & \multicolumn{1}{c}{\multirow{2}{*}{\textbf{\begin{tabular}[c]{@{}c@{}}Err. of \\ Hig. App.\end{tabular}}}} & \multicolumn{1}{c}{\multirow{2}{*}{\textbf{\begin{tabular}[c]{@{}c@{}}Err. of\\ FPCC\end{tabular}}}} & \multicolumn{1}{c}{\multirow{2}{*}{\textbf{\begin{tabular}[c]{@{}c@{}}Function\\ Index\end{tabular}}}} & \multicolumn{1}{c}{\multirow{2}{*}{\textbf{\begin{tabular}[c]{@{}c@{}}Err. of\\ Our App.\end{tabular}}}} & \multicolumn{1}{c}{\multirow{2}{*}{\textbf{\begin{tabular}[c]{@{}c@{}}Err. of \\ Hig. App.\end{tabular}}}} & \multicolumn{1}{c}{\multirow{2}{*}{\textbf{\begin{tabular}[c]{@{}c@{}}Err. of\\ FPCC\end{tabular}}}}                   \\ \multicolumn{1}{c}{}                                                                                   & \multicolumn{1}{c}{} & \multicolumn{1}{c}{}  & \multicolumn{1}{c}{}                                                                                 & \multicolumn{1}{c}{}                                                                                   & \multicolumn{1}{c}{}                                                                                     & \multicolumn{1}{c}{}                                                                                       & \multicolumn{1}{c}{}    & \multicolumn{1}{c}{}    & \multicolumn{1}{c}{}    & \multicolumn{1}{c}{}    & \multicolumn{1}{c}{}       \\                                                                                                  \cmidrule(lr){1-4} \cmidrule(lr){5-8} \cmidrule(lr){9-12}                                                                                 
ATOM\_1  & 1.07E+05          & 1.39E+04 & 7.19E+01          & ATOM\_59  & 7.32E-01 & 7.03E+00  & 2.04E+00          & ATOM\_117 & 5.00E-01 & 1.51E-02 & \textbf{1.32E-16} \\
ATOM\_2  & 9.82E-01          & 3.96E+02 & 4.66E+00          & ATOM\_60  & 2.17E+00 & 1.37E-01  & 1.28E+00          & ATOM\_118 & 1.00E+00 & 8.66E-01 & 2.50E+27          \\
ATOM\_3  & 1.54E+00          & 3.00E+00 & 1.74E+01          & ATOM\_61  & 2.61E-02 & 1.06E-02  & \textbf{1.52E-16} & ATOM\_119 & 1.00E+00 & 1.00E+00 & 1.48E+27          \\
ATOM\_4  & 5.23E-02          & 1.60E-01 & 3.81E-02          & ATOM\_62  & 2.94E+42 & 1.00E+00  & 2.93E+06          & ATOM\_120 & 1.00E+00 & 6.16E-01 & 2.16E+27          \\
ATOM\_5  & 5.81E+51          & 1.47E+00 & 1.24E+02          & ATOM\_63  & 4.97E+09 & 8.30E-01  & 3.09E+02          & ATOM\_121 & 1.00E+00 & 5.03E-01 & 1.73E+00          \\
ATOM\_6  & 7.03E-02          & 6.57E+00 & 7.03E+01          & ATOM\_64  & 1.61E+02 & 1.44E+00  & 8.78E+00          & ATOM\_122 & 2.50E-01 & 2.85E-02 & 2.42E-01          \\
ATOM\_7  & 2.28E+05          & 1.00E+00 & 4.04E+06          & ATOM\_65  & 5.88E-02 & 8.32E-03  & 4.32E-01          & ATOM\_123 & 1.10E-02 & 2.84E-03 & \textbf{6.04E-18} \\
ATOM\_8  & 1.00E+00          & 2.99E-01 & 1.45E+00          & ATOM\_66  & 1.54E+00 & 2.09E+00  & 9.31E+00          & ATOM\_124 & 5.39E-03 & 2.30E-03 & \textbf{1.24E-16} \\
ATOM\_9  & 2.60E-01          & 4.38E+00 & 1.57E+00          & ATOM\_67  & 3.26E-02 & 1.38E-02  & 3.61E-02          & ATOM\_125 & 1.00E+00 & 7.13E-01 & 2.20E+00          \\
ATOM\_10 & 5.19E-03          & 7.88E-03 & 5.74E-03          & ATOM\_68  & 1.76E+00 & 2.49E+00  & \textbf{5.98E-06} & ATOM\_126 & 1.00E+00 & 1.10E-01 & \textbf{2.22E-16} \\
ATOM\_11 & 2.57E+74          & 1.11E+00 & 2.45E+07          & ATOM\_69  & 1.49E-02 & 5.00E+00  & 1.77E+01          & ATOM\_127 & 5.00E-01 & 1.10E-01 & \textbf{1.11E-16} \\
ATOM\_12 & 2.58E+09          & 1.00E+00 & 7.54E+06          & ATOM\_70  & 1.91E+00 & 1.24E-01  & 2.00E+00          & ATOM\_128 & 5.18E-01 & 1.31E-01 & 2.68E-01          \\
ATOM\_13 & 1.29E+06          & 1.00E+00 & 2.23E+06          & ATOM\_71  & 1.07E+05 & 1.81E+04  & 4.22E+01          & ATOM\_129 & 4.17E-02 & 1.18E-01 & 2.25E+00          \\
ATOM\_14 & 9.93E-01          & 5.43E+09 & 6.70E+00          & ATOM\_72  & 3.11E+00 & 3.43E+01  & 1.18E+01          & ATOM\_130 & 3.33E-01 & 9.93E-02 & \textbf{1.11E-16} \\
ATOM\_15 & 2.42E+07          & 8.85E+00 & 8.09E+00          & ATOM\_73  & 1.54E+00 & 1.87E+00  & 9.31E+00          & ATOM\_131 & 5.00E-01 & 9.93E-02 & \textbf{2.22E-16} \\
ATOM\_16 & 1.00E+00          & 6.04E+03 & 1.18E+01          & ATOM\_74  & 4.50E-03 & 1.02E-03  & 4.67E-03          & ATOM\_132 & 3.33E-01 & 9.93E-02 & \textbf{1.11E-16} \\
ATOM\_17 & 1.09E+00          & 5.06E+01 & 1.09E+01          & ATOM\_75  & 1.56E+60 & 3.62E-03  & 4.73E+03          & ATOM\_133 & 5.56E-01 & 2.68E-01 & \textbf{7.92E-15} \\
ATOM\_18 & 1.85E+00          & 9.04E-03 & 5.83E+00          & ATOM\_76  & 2.81E+02 & 1.40E+01  & 5.63E+01          & ATOM\_134 & 7.50E-01 & 1.89E-01 & \textbf{1.48E-16} \\
ATOM\_19 & 3.39E-02          & 6.22E-02 & 6.42E-02          & ATOM\_77  & 3.57E+07 & 1.00E+00  & 3.02E+06          & ATOM\_135 & 5.56E-01 & 2.68E-01 & \textbf{1.48E-16} \\
ATOM\_20 & 7.06E+72          & 2.19E+00 & 4.03E+05          & ATOM\_78  & 6.07E-01 & 4.43E-02  & 6.93E-01          & ATOM\_136 & 2.13E-01 & 1.52E-02 & 7.22E-02          \\
ATOM\_21 & 2.81E+02          & 1.13E+00 & 1.66E+02          & ATOM\_79  & 5.82E-01 & 2.44E-01  & 8.37E-01          & ATOM\_137 & 8.70E-02 & 3.47E-02 & 4.48E+00          \\
ATOM\_22 & 4.59E+08          & 1.00E+00 & 1.66E+07          & ATOM\_80  & 6.48E+01 & 9.21E-01  & 1.39E+02          & ATOM\_138 & 6.51E-03 & 1.12E-03 & 6.77E-03          \\
ATOM\_23 & 1.00E+00          & 2.19E-01 & 1.45E+00          & ATOM\_81  & 3.56E+00 & 4.77E-01  & 1.15E-03          & ATOM\_139 & 7.68E+11 & 1.00E+00 & 2.89E+07          \\
ATOM\_24 & 3.59E-01          & 1.54E+01 & 1.39E+01          & ATOM\_82  & 1.00E+00 & 4.29E-01  & 1.47E+00          & ATOM\_140 & 7.54E+07 & 1.00E+00 & 4.61E+06          \\
ATOM\_25 & 5.74E-02          & 7.13E-02 & 3.98E-02          & ATOM\_83  & 1.00E+00 & 5.55E-01  & 2.70E+00          & ATOM\_141 & 9.93E-01 & 3.13E+09 & 2.94E+00          \\
ATOM\_26 & 4.62E+20          & 1.00E+00 & 3.18E+08          & ATOM\_84  & 1.19E-02 & 1.61E-02  & 1.36E-01          & ATOM\_142 & 2.42E+07 & 9.25E+00 & 4.09E+00          \\
ATOM\_27 & 1.68E+05          & 1.00E+00 & 1.10E+07          & ATOM\_85  & 5.00E-01 & 2.38E+00  & 6.68E+00          & ATOM\_143 & 6.86E+02 & 9.99E+04 & 7.94E+00          \\
ATOM\_28 & 9.93E-01          & 2.47E+09 & 7.04E+00          & ATOM\_86  & 1.19E-02 & 1.61E-02  & 1.36E-01          & ATOM\_144 & 1.09E+00 & 3.90E+01 & 3.77E+00          \\
ATOM\_29 & 3.16E+10          & 6.84E+00 & 8.48E+02          & ATOM\_87  & 1.00E+00 & 3.01E+212 & 1.25E+00          & ATOM\_145 & 1.54E+00 & 1.54E+00 & 9.31E+00          \\
ATOM\_30 & 6.79E-02          & 8.05E-03 & 7.12E-02          & ATOM\_88  & 1.00E+00 & 1.00E+00  & 1.43E+00          & ATOM\_146 & 5.43E-02 & 2.52E-02 & 6.22E-02          \\
ATOM\_31 & 5.00E-01          & 2.10E-01 & 5.62E-01          & ATOM\_89  & 1.72E-02 & 5.11E-03  & 3.63E-01          & ATOM\_147 & 1.95E+13 & 5.74E-02 & \textbf{7.61E-05} \\
ATOM\_32 & 1.53E-01          & 6.27E-02 & 1.45E-01          & ATOM\_90  & 4.55E-02 & 2.16E-01  & 9.57E-01          & ATOM\_148 & 2.81E+02 & 1.69E-02 & 5.63E+01          \\
ATOM\_33 & 1.43E-01          & 1.47E-02 & 1.61E-01          & ATOM\_91  & 1.25E-01 & 1.19E-02  & 7.29E-01          & ATOM\_149 & 1.53E+05 & 1.00E+00 & 2.19E+07          \\
ATOM\_34 & 9.09E-02          & 8.53E-03 & 1.02E-01          & ATOM\_92  & 2.17E+00 & 1.38E-01  & 1.28E+00          & ATOM\_150 & 8.70E-02 & 1.00E+00 & 7.04E+00          \\
ATOM\_35 & 1.67E-01          & 6.53E-01 & 4.70E+00          & ATOM\_93  & 3.76E-02 & 4.38E-03  & 4.06E-02          & ATOM\_151 & 2.18E+16 & 1.00E+00 & 1.01E+08          \\
ATOM\_36 & 1.90E+00          & 6.23E-03 & \textbf{2.54E-16} & ATOM\_94  & 4.83E+11 & 1.00E+00  & 1.82E+07          & ATOM\_152 & 2.92E+05 & 1.00E+00 & 2.58E+07          \\
ATOM\_37 & \textbf{7.47E-04} & 1.17E-03 & 1.06E-02          & ATOM\_95  & 2.29E+06 & 1.00E+00  & 1.40E+07          & ATOM\_153 & 7.07E+05 & 1.00E+00 & 1.74E+06          \\
ATOM\_38 & 1.43E-01          & 3.37E-03 & 2.34E-01          & ATOM\_96  & 2.98E+03 & 4.96E+04  & 5.31E+00          & ATOM\_154 & 1.01E+03 & 1.00E+00 & 1.59E+01          \\
ATOM\_39 & 7.69E-02          & 2.93E-01 & 9.30E-01          & ATOM\_97  & 7.32E-01 & 1.03E+01  & 2.04E+00          & ATOM\_155 & 2.42E+07 & 1.00E+00 & 6.42E+00          \\
ATOM\_40 & 2.46E-03          & 4.30E-02 & 4.65E-02          & ATOM\_98  & 2.17E+00 & 5.82E-02  & 1.28E+00          & ATOM\_156 & 5.00E-01 & 1.62E-01 & 3.50E-02          \\
ATOM\_41 & 2.50E-01          & 2.52E+00 & 8.55E+00          & ATOM\_99  & 3.18E-02 & 2.18E-02  & 7.49E-02          & ATOM\_157 & 1.00E-01 & 1.50E-02 & 1.18E-03          \\
ATOM\_42 & 1.59E-01          & 1.77E-02 & \textbf{1.41E-16} & ATOM\_100 & 1.23E+19 & 1.00E+00  & 3.87E+06          & ATOM\_158 & 3.33E-01 & 6.58E-02 & \textbf{2.57E-14} \\
ATOM\_43 & 2.00E-02          & 2.38E-01 & 2.13E-02          & ATOM\_101 & 2.77E+03 & 1.44E+00  & 1.55E+02          & ATOM\_159 & 1.00E+00 & 1.13E-01 & 2.00E-01          \\
ATOM\_44 & 1.61E-02          & 8.79E-02 & 3.61E-02          & ATOM\_102 & 1.61E+02 & 8.16E-01  & 8.78E+00          & ATOM\_160 & 1.43E-01 & 5.08E-02 & \textbf{2.90E-16} \\
ATOM\_45 & 7.81E+03          & 1.35E+03 & 8.72E+00          & ATOM\_103 & 5.00E-01 & 1.53E+00  & 2.18E+00          & ATOM\_161 & 6.25E-02 & 2.08E-02 & \textbf{2.90E-16} \\
ATOM\_46 & 9.82E-01          & 3.94E+01 & 1.89E+00          & ATOM\_104 & 1.32E-02 & 1.56E-01  & 1.40E+00          & ATOM\_162 & 2.50E-01 & 5.08E-02 & \textbf{3.45E-15} \\
ATOM\_47 & 1.54E+00          & 1.54E+00 & 9.31E+00          & ATOM\_105 & 1.00E+00 & 1.10E-01  & \textbf{2.22E-16} & ATOM\_163 & 3.37E-01 & 5.05E-02 & 3.91E-03          \\
ATOM\_48 & 1.62E-02          & 3.51E-03 & 1.72E-02          & ATOM\_106 & 1.00E+00 & 1.10E-01  & \textbf{2.22E-16} & ATOM\_164 & 5.00E-01 & 9.93E-02 & \textbf{2.90E-16} \\
ATOM\_49 & 9.52E+55          & 3.98E+00 & 2.79E+02          & ATOM\_107 & 3.33E-01 & 4.84E-02  & \textbf{2.22E-16} & ATOM\_165 & 5.00E-01 & 9.93E-02 & \textbf{3.45E-15} \\
ATOM\_50 & 1.24E+08          & 2.19E+00 & 1.13E+02          & ATOM\_108 & 2.00E-01 & 1.10E-01  & \textbf{1.11E-16} & ATOM\_166 & 2.50E-01 & 5.08E-02 & \textbf{3.45E-15} \\
ATOM\_51 & 3.95E+05          & 1.00E+00 & 3.71E+06          & ATOM\_109 & 5.00E-01 & 1.12E-01  & \textbf{2.22E-16} & ATOM\_167 & 3.37E-01 & 5.05E-02 & 3.91E-03          \\
ATOM\_52 & 1.43E-01          & 5.51E-02 & 1.44E+00          & ATOM\_110 & 2.50E-01 & 1.31E-02  & \textbf{1.11E-16} & ATOM\_168 & 5.50E-01 & 1.01E-01 & 9.88E-02          \\
ATOM\_53 & 2.17E+00          & 1.37E-01 & 1.28E+00          & ATOM\_111 & 1.11E-01 & 2.50E+00  & 2.07E+00          & ATOM\_169 & 6.03E-01 & 6.17E-02 & 9.47E-02          \\
ATOM\_54 & 2.48E-02          & 1.98E-03 & 2.68E-02          & ATOM\_112 & 1.03E-02 & 2.35E-02  & 5.92E-03          & ATOM\_170 & 2.03E-01 & 1.63E-02 & 2.35E-03          \\
ATOM\_55 & 3.65E+19          & 1.00E+00 & 1.15E+07          & ATOM\_113 & 5.00E-01 & 2.96E-01  & \textbf{9.46E-14} & ATOM\_171 & 2.03E-01 & 1.63E-02 & 2.35E-03          \\
ATOM\_56 & 2.44E+12          & 1.00E+00 & 2.31E+07          & ATOM\_114 & 2.86E-01 & 8.09E-02  & \textbf{2.05E-16} & ATOM\_172 & 3.37E-01 & 5.05E-02 & 3.91E-03          \\
ATOM\_57 & 9.54E+05          & 1.00E+00 & \textbf{1.44E-16} & ATOM\_115 & 2.00E-01 & 1.33E-02  & 2.50E-01          & HSED\_1   & 7.25E-03 & 3.58E-03 & 9.23E-03          \\
ATOM\_58 & 4.07E+06          & 2.51E+09 & \textbf{4.69E-32} & ATOM\_116 & 7.14E-02 & 2.86E-02  & \textbf{1.11E-16} & HSED\_2   & 7.25E-03 & 3.58E-03 & 9.23E-03  \\
\cmidrule(lr){1-4} \cmidrule(lr){5-8} \cmidrule(lr){9-12} 
\end{tabular}
\vspace{-0.4cm}
\end{table*}

\textbf{Errors produced by our approach exhibit strong monotonic trends near the point with local maximum floating-point error.} After filtering out cases with inputs exceeding \(10^9\), we have 79 significant errors, corresponding to 158 sides. For certain sides, the function values are computationally infeasible, and these are excluded from further analysis. Among the remaining valid cases, both PI-detector and the high-precision approach demonstrate consistent monotonic trends, with all p-values falling well below the 0.05 significance threshold. Since FPCC can not detect some significant errors, 28 out of the 79 cases illustrate the wrong monotonicity. The Mann-Kendall statistic \(S\) for our approach yields average values of 499,373.52 (increasing trend) and -500,613.90 (decreasing trend), reflecting robust monotonicity. By comparison, the high-precision approach produces substantially lower magnitudes (227,762.44 and -239,604.52, respectively), highlighting the PI-detector’s ability to mitigate the instability inherent in the high-precision approach (see Figure~\ref{rq1}). When excluding the 28 erroneous FPCC cases, its Mann-Kendall statistic improves to 495,988.05 (increasing) and -478,551.53 (decreasing), suggesting relatively strong—though still inferior to the PI-detector—monotonic trends.

\begin{figure}[htbp]
    \centering
    \includegraphics[width=0.48\textwidth]{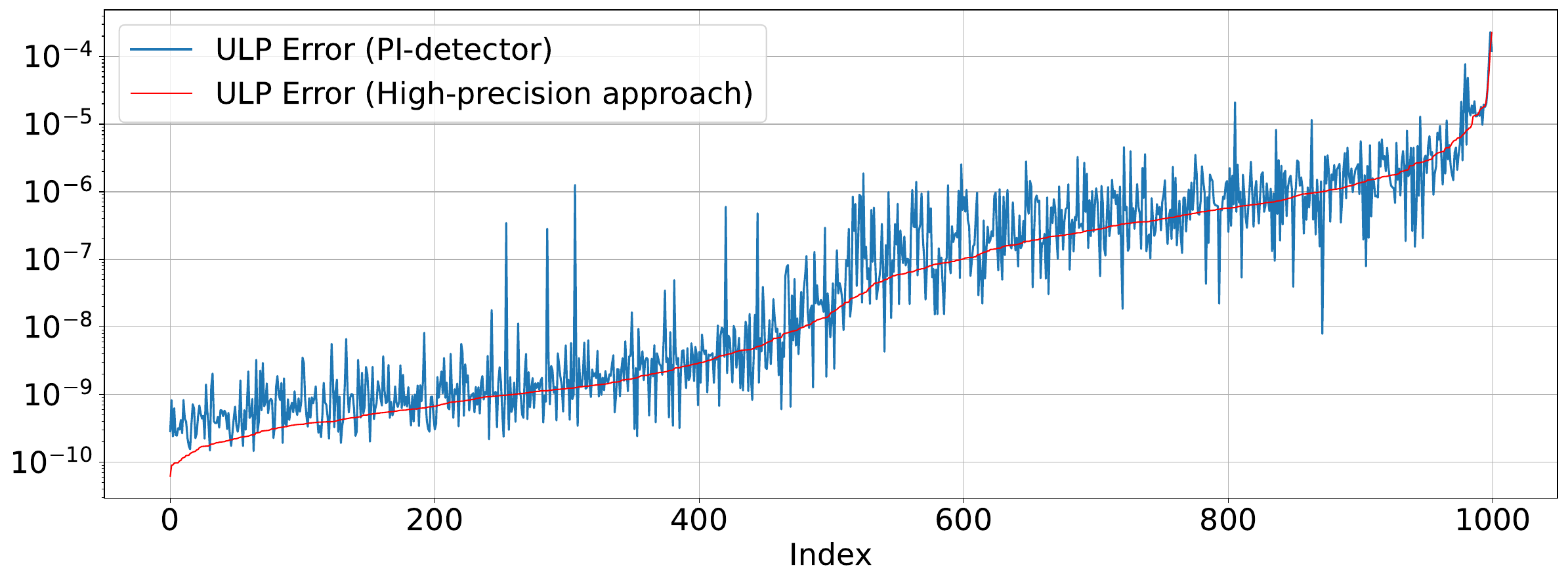}
     \vspace{-0.2cm}
    \caption{Errors calculated by PI-detector and high-precision approach.}
    \label{correlation_matrix}
\end{figure}

\textbf{PI-detector demonstrates exceptional effectiveness in complex floating-point programs.} The Pearson correlation and Spearman correlation scores between the errors computed by our approach and the high-precision approach are 0.8179 and 0.9243, respectively, as shown in Figure~\ref{correlation_matrix}. This strong correlation demonstrates the accuracy of our approach in quantifying errors within complex floating-point programs. Notably, the matrix condition numbers also illustrate a similar trend with errors computed by PI-detector, which is consistent with the theory of numerical linear algebra~\cite{ford2014numerical}. This remarkable performance highlights PI-detector’s robustness when confronted with complex scenarios.

\vspace{-0.9cm}
\findingboxx{The floating-point errors calculated by our approach is reliable, and can be used to detect significant errors. Notably, FPCC cannot be used with our complex datasets because it does not support multi-input data, as specified in its replication package. In contrast, PI-detector successfully processes complex floating-point programs, indicating both robust performance and strong generalizability.}

\subsection*{RQ2 (addressing limitation 2 (prolonged execution time) and mitigating the slow execution of FPCC.): Is PI-detector capable of computing errors efficiently?}

\noindent\textit{\textbf{Motivation.}} One limitation of the high-precision approach is its prolonged execution time, which substantially degrades the performance of tools that rely on floating-point error computation. Although the state-of-the-art method, FPCC, enhances execution efficiency by estimating the overall condition number of the program—referred to as the chain condition number—the computation remains time-consuming. To assess whether PI-detector can address this efficiency issue, we compare the runtime of our approach with that of both the high-precision approach and FPCC.

\noindent\textit{\textbf{Approach.}} To evaluate whether PI-detector can reduce execution time compared to the high-precision approach and FPCC, we involve the following steps. First, we calculate errors of 2,000 surrounding points for all 174 cases. We execute these calculations using the three approaches to create a comprehensive dataset for comparison. Afterward, we record the total execution time for each case. In addition to the ATOMU and HSED datasets, we extend our evaluation to include complex programs. We record the execution time of the 1,000 linear system-solving program cases (cf. Section~\ref{rques1}).

\noindent\textit{\textbf{Results. }}\textbf{PI-detector can save excessive time compared to the high-precision approach and FPCC for both ATOMU and HSED datasets.} The execution time of PI-detector comprises the combined runtime of both the original GSL program and its perturbed version. As shown in Table~\ref{time_table}, PI-detector achieves an average runtime of just 3.35 ms across all 174 test cases. In contrast, the average execution time of the high-precision approach is 1,366.23 ms, reflecting a high rise of about 407 times compared to our approach. The average runtime of FPCC is 34.24 ms, indicating that FPCC requires more than ten times the computational time of our approach. While FPCC shows superior speed in certain cases, these are limited to very short-running, trivial functions. Such dramatic reductions underscore the PI-detector’s efficiency in reducing execution time while still producing reliable results.

\noindent\textbf{PI-detector performs well for complex programs with long execution time.} As for the 1,000 3,000-dimensional linear system-solving programs, the total execution time of the original program is 0.6414 hours, rising to 0.9893 hours for the perturbed version. In contrast, the total runtime of the high-precision program is 73.6650 hours, making it about 44 times longer than the execution time of PI-detector. This disparity illustrates PI-detector's capacity to effectively handle long-runtime floating-point programs. For example, consider a numerical simulation program that typically requires a full day to complete~\cite{he2020testing}. Using a high-precision program to evaluate a specific input for significant errors could extend the runtime to approximately 100 days. PI-detector, on the other hand, offers a practical solution, significantly reducing this time while still ensuring accurate error calculation.

\begin{table*}[t]
\caption{Runtime of floating-point error calculations by using PI-detector, FPCC, and high-precision approach (Note: \textbf{App.} = Approach, \textbf{Hig.} = High-precision).} 
\vspace{-.2cm}
\centering
\footnotesize
\label{time_table}
\resizebox{\textwidth}{!}{
\setlength{\tabcolsep}{3pt}
\begin{tabular}{lrrrlrrrlrrrlrrrlrrr}
\cmidrule(lr){1-4} \cmidrule(lr){5-8} \cmidrule(lr){9-12} \cmidrule(lr){13-16} \cmidrule(lr){17-20}
\multicolumn{1}{c}{\multirow{2}{*}{\textbf{\begin{tabular}[c]{@{}c@{}}Function\\ Index\end{tabular}}}} &  \multicolumn{1}{c}{\multirow{2}{*}{\textbf{\begin{tabular}[c]{@{}c@{}}PI-D\\ (ms)\end{tabular}}}} &\multicolumn{1}{c}{\multirow{2}{*}{\textbf{\begin{tabular}[c]{@{}c@{}}FPCC\\ (ms)\end{tabular}}}} &\multicolumn{1}{c}{\multirow{2}{*}{\textbf{\begin{tabular}[c]{@{}c@{}}Hig.\\ (ms)\end{tabular}}}}& \multicolumn{1}{c}{\multirow{2}{*}{\textbf{\begin{tabular}[c]{@{}c@{}}Function\\ Index\end{tabular}}}} &  \multicolumn{1}{c}{\multirow{2}{*}{\textbf{\begin{tabular}[c]{@{}c@{}}PI-D\\ (ms)\end{tabular}}}} &\multicolumn{1}{c}{\multirow{2}{*}{\textbf{\begin{tabular}[c]{@{}c@{}}FPCC\\ (ms)\end{tabular}}}} &\multicolumn{1}{c}{\multirow{2}{*}{\textbf{\begin{tabular}[c]{@{}c@{}}Hig.\\ (ms)\end{tabular}}}}&\multicolumn{1}{c}{\multirow{2}{*}{\textbf{\begin{tabular}[c]{@{}c@{}}Function\\ Index\end{tabular}}}} &  \multicolumn{1}{c}{\multirow{2}{*}{\textbf{\begin{tabular}[c]{@{}c@{}}PI-D\\ (ms)\end{tabular}}}} &\multicolumn{1}{c}{\multirow{2}{*}{\textbf{\begin{tabular}[c]{@{}c@{}}FPCC\\ (ms)\end{tabular}}}} &\multicolumn{1}{c}{\multirow{2}{*}{\textbf{\begin{tabular}[c]{@{}c@{}}Hig.\\  (ms)\end{tabular}}}}  & \multicolumn{1}{c}{\multirow{2}{*}{\textbf{\begin{tabular}[c]{@{}c@{}}Function\\ Index\end{tabular}}}} &  \multicolumn{1}{c}{\multirow{2}{*}{\textbf{\begin{tabular}[c]{@{}c@{}}PI-D\\ (ms)\end{tabular}}}} &\multicolumn{1}{c}{\multirow{2}{*}{\textbf{\begin{tabular}[c]{@{}c@{}}FPCC\\ (ms)\end{tabular}}}} &\multicolumn{1}{c}{\multirow{2}{*}{\textbf{\begin{tabular}[c]{@{}c@{}}Hig.\\  (ms)\end{tabular}}}}& \multicolumn{1}{c}{\multirow{2}{*}{\textbf{\begin{tabular}[c]{@{}c@{}}Function\\ Index\end{tabular}}}} &  \multicolumn{1}{c}{\multirow{2}{*}{\textbf{\begin{tabular}[c]{@{}c@{}}PI-D\\ (ms)\end{tabular}}}} &\multicolumn{1}{c}{\multirow{2}{*}{\textbf{\begin{tabular}[c]{@{}c@{}}FPCC\\ (ms)\end{tabular}}}} &\multicolumn{1}{c}{\multirow{2}{*}{\textbf{\begin{tabular}[c]{@{}c@{}}Hig.\\  (ms)\end{tabular}}}}           \\ \multicolumn{1}{c}{}                                                                                   & \multicolumn{1}{c}{} & \multicolumn{1}{c}{}  & \multicolumn{1}{c}{}                                                                                 & \multicolumn{1}{c}{}                                                                                   & \multicolumn{1}{c}{}                                                                                     & \multicolumn{1}{c}{}                                                                                       & \multicolumn{1}{c}{}    & \multicolumn{1}{c}{}    & \multicolumn{1}{c}{}    & \multicolumn{1}{c}{}    & \multicolumn{1}{c}{}   & \multicolumn{1}{c}{} & \multicolumn{1}{c}{} & \multicolumn{1}{c}{} & \multicolumn{1}{c}{} & \multicolumn{1}{c}{} & \multicolumn{1}{c}{} & \multicolumn{1}{c}{} & \multicolumn{1}{c}{}     \\                                                                                                  \cmidrule(lr){1-4} \cmidrule(lr){5-8} \cmidrule(lr){9-12} \cmidrule(lr){13-16} \cmidrule(lr){17-20}                                                                                

ATOM\_1  & 8           & 56   & 816   & ATOM\_36 & 6 & 58 & 526   & ATOM\_71  & 3 & 36  & 39    & ATOM\_106 & 1  & 3   & 31      & ATOM\_141 & 2 & 13  & 16    \\
ATOM\_2  & 8           & 62   & 773   & ATOM\_37 & 6 & 58 & 28    & ATOM\_72  & 3 & 32  & 40    & ATOM\_107 & 1  & 3   & 34      & ATOM\_142 & 2 & 17  & 16    \\
ATOM\_3  & 8           & 55   & 777   & ATOM\_38 & 2 & 12 & 25    & ATOM\_73  & 3 & 35  & 40    & ATOM\_108 & 1  & 4   & 33      & ATOM\_143 & 1 & 13  & 15    \\
ATOM\_4  & 9           & 83   & 757   & ATOM\_39 & 3 & 23 & 443   & ATOM\_74  & 3 & 78  & 42    & ATOM\_109 & 1  & 3   & 33      & ATOM\_144 & 1 & 12  & 15    \\
ATOM\_5  & 9           & 60   & 768   & ATOM\_40 & 6 & 47 & 1,297 & ATOM\_75  & 3 & 41  & 42    & ATOM\_110 & 1  & 4   & 32      & ATOM\_145 & 2 & 13  & 15    \\
ATOM\_6  & 9           & 56   & 762   & ATOM\_41 & 3 & 25 & 1,113 & ATOM\_76  & 3 & 28  & 39    & ATOM\_111 & 3  & 6   & 32      & ATOM\_146 & 2 & 35  & 16    \\
ATOM\_7  & 8           & 57   & 753   & ATOM\_42 & 1 & 0  & 19    & ATOM\_77  & 3 & 30  & 42    & ATOM\_112 & 11 & 76  & 34      & ATOM\_147 & 2 & 13  & 15    \\
ATOM\_8  & 6           & 38   & 212   & ATOM\_43 & 1 & 1  & 30    & ATOM\_78  & 3 & 16  & 617   & ATOM\_113 & 1  & 6   & 52      & ATOM\_148 & 2 & 15  & 15    \\
ATOM\_9  & 8           & 70   & 755   & ATOM\_44 & 1 & 2  & 42    & ATOM\_79  & 3 & 32  & 2,270 & ATOM\_114 & 2  & 4   & 24      & ATOM\_149 & 2 & 12  & 15    \\
ATOM\_10 & 8           & 83   & 742   & ATOM\_45 & 2 & 12 & 18    & ATOM\_80  & 3 & 30  & 2,302 & ATOM\_115 & 2  & 5   & 25      & ATOM\_150 & 2 & 13  & 14    \\
ATOM\_11 & 8           & 58   & 773   & ATOM\_46 & 2 & 15 & 18    & ATOM\_81  & 1 & 2   & 438   & ATOM\_116 & 2  & 54  & 27      & ATOM\_151 & 2 & 13  & 14    \\
ATOM\_12 & 8           & 55   & 756   & ATOM\_47 & 2 & 13 & 21    & ATOM\_82  & 3 & 14  & 4,130 & ATOM\_117 & 3  & 54  & 40      & ATOM\_152 & 2 & 14  & 15    \\
ATOM\_13 & 8           & 71   & 791   & ATOM\_48 & 2 & 36 & 20    & ATOM\_83  & 2 & 10  & 433   & ATOM\_118 & 2  & 7   & 50      & ATOM\_153 & 2 & 26  & 27    \\
ATOM\_14 & 8           & 56   & 760   & ATOM\_49 & 2 & 15 & 21    & ATOM\_84  & 2 & 16  & 41    & ATOM\_119 & 2  & 6   & 58      & ATOM\_154 & 2 & 13  & 14    \\
ATOM\_15 & 8           & 59   & 762   & ATOM\_50 & 2 & 25 & 19    & ATOM\_85  & 3 & 23  & 69    & ATOM\_120 & 2  & 6   & 48      & ATOM\_155 & 2 & 12  & 14    \\
ATOM\_16 & 8           & 60   & 817   & ATOM\_51 & 2 & 20 & 18    & ATOM\_86  & 2 & 15  & 53    & ATOM\_121 & 1  & 1   & 42      & ATOM\_156 & 1 & 3   & 23    \\
ATOM\_17 & 8           & 59   & 812   & ATOM\_52 & 3 & 30 & 31    & ATOM\_87  & 4 & 38  & 50    & ATOM\_122 & 1  & 1   & 46      & ATOM\_157 & 1 & 30  & 479   \\
ATOM\_18 & 8           & 68   & 810   & ATOM\_53 & 3 & 49 & 30    & ATOM\_88  & 4 & 42  & 46    & ATOM\_123 & 1  & 1   & 1,392   & ATOM\_158 & 1 & 33  & 479   \\
ATOM\_19 & 8           & 96   & 790   & ATOM\_54 & 3 & 78 & 33    & ATOM\_89  & 2 & 17  & 27    & ATOM\_124 & 1  & 1   & 1,356   & ATOM\_159 & 4 & 102 & 477   \\
ATOM\_20 & 8           & 61   & 808   & ATOM\_55 & 3 & 27 & 34    & ATOM\_90  & 2 & 16  & 36    & ATOM\_125 & 1  & 1   & 16,151  & ATOM\_160 & 3 & 20  & 680   \\
ATOM\_21 & 8           & 57   & 807   & ATOM\_56 & 3 & 33 & 31    & ATOM\_91  & 4 & 29  & 66    & ATOM\_126 & 1  & 0   & 15      & ATOM\_161 & 3 & 20  & 665   \\
ATOM\_22 & 8           & 63   & 854   & ATOM\_57 & 3 & 63 & 29    & ATOM\_92  & 3 & 51  & 28    & ATOM\_127 & 1  & 0   & 16      & ATOM\_162 & 3 & 20  & 666   \\
ATOM\_23 & 7           & 38   & 258   & ATOM\_58 & 3 & 36 & 30    & ATOM\_93  & 3 & 78  & 31    & ATOM\_128 & 3  & 108 & 62      & ATOM\_163 & 1 & 29  & 253   \\
ATOM\_24 & 9           & 69   & 833   & ATOM\_59 & 3 & 29 & 30    & ATOM\_94  & 3 & 29  & 28    & ATOM\_129 & 3  & 16  & 73      & ATOM\_164 & 3 & 20  & 665   \\
ATOM\_25 & 8           & 82   & 808   & ATOM\_60 & 3 & 46 & 37    & ATOM\_95  & 3 & 39  & 28    & ATOM\_130 & 3  & 62  & 69      & ATOM\_165 & 3 & 21  & 666   \\
ATOM\_26 & 8           & 57   & 813   & ATOM\_61 & 3 & 77 & 44    & ATOM\_96  & 3 & 30  & 28    & ATOM\_131 & 3  & 24  & 74      & ATOM\_166 & 3 & 20  & 667   \\
ATOM\_27 & 8           & 57   & 2,308 & ATOM\_62 & 3 & 36 & 30    & ATOM\_97  & 3 & 29  & 28    & ATOM\_132 & 3  & 22  & 70      & ATOM\_167 & 1 & 31  & 274   \\
ATOM\_28 & 8           & 56   & 7,304 & ATOM\_63 & 3 & 23 & 29    & ATOM\_98  & 3 & 54  & 28    & ATOM\_133 & 3  & 13  & 434     & ATOM\_168 & 4 & 106 & 538   \\
ATOM\_29 & 8           & 57   & 802   & ATOM\_64 & 3 & 30 & 30    & ATOM\_99  & 3 & 80  & 31    & ATOM\_134 & 3  & 12  & 419     & ATOM\_169 & 4 & 106 & 527   \\
ATOM\_30 & 6           & 80   & 233   & ATOM\_65 & 3 & 26 & 43    & ATOM\_100 & 3 & 28  & 27    & ATOM\_135 & 3  & 12  & 415     & ATOM\_170 & 1 & 30  & 274   \\
ATOM\_31 & 5           & 24   & 252   & ATOM\_66 & 3 & 37 & 40    & ATOM\_101 & 3 & 32  & 29    & ATOM\_136 & 3  & 52  & 164,797 & ATOM\_171 & 1 & 30  & 271   \\
ATOM\_32 & 4           & 18   & 1,692 & ATOM\_67 & 3 & 78 & 43    & ATOM\_102 & 3 & 33  & 31    & ATOM\_137 & 2  & 13  & 16      & ATOM\_172 & 1 & 30  & 271   \\
ATOM\_33 & 5           & 25   & 293   & ATOM\_68 & 3 & 31 & 40    & ATOM\_103 & 2 & 13  & 27    & ATOM\_138 & 2  & 34  & 17      & HSED\_1   & 1 & 0   & 10    \\
ATOM\_34 & 5           & 26   & 288   & ATOM\_69 & 3 & 30 & 39    & ATOM\_104 & 8 & 185 & 39    & ATOM\_139 & 2  & 15  & 15      & HSED\_2   & 1 & 0   & 19    \\
ATOM\_35 & 2           & 11   & 23    & ATOM\_70 & 3 & 64 & 39    & ATOM\_105 & 1 & 3   & 33    & ATOM\_140 & 2  & 13  & 16      & Average   & 3.35 & 34.24  & 1,366.23 \\
 
\cmidrule(lr){1-4} \cmidrule(lr){5-8} \cmidrule(lr){9-12} \cmidrule(lr){13-16} \cmidrule(lr){17-20}
\end{tabular}
}
\vspace{-.4cm}
\end{table*}

\findingboxx{PI-detector significantly reduces execution time compared to FPCC and the high-precision approach while maintaining reliable error computation. Notably, our approach maintains high efficiency even in complex programs, demonstrating its broad application potential. FPCC fails to process our complex datasets, demonstrating that it is not only slower than our approach but also has limited applicability.} 


\section{Threats to Validity}
\label{threat}
In this section, we discuss the threats to the validity of this work.

\noindent \textbf{External validity.} Our evaluation primarily focuses on C programs, which raises concerns about the generalization of PI-detector across different programming languages. PI-detector may not perform effectively for floating-point programs in high-level languages such as Python and Java. However, it is important to note that PI-detector is built upon LLVM pass, enabling its application to any language that can be converted to LLVM Intermediate Representation (IR).

\noindent \textbf{Internal validity.} We ensure internal validity by only modifying the operands through perturbation injections, without modifying any other parts of the program logic or execution environment. Therefore, we can be confident that the observed changes in program output are solely due to the introduced perturbations, not other external factors.

\noindent \textbf{Construct validity.} If we only evaluate whether our approach can detect significant errors, the apparent effectiveness might stem from systematic overestimation in our error calculations. To ensure construct validity, we measure True Positives (TP), False Positives (FP), False Negatives (FN), and True Negatives (TN) in our experiments. We also assess the monotonic trend of floating-point errors near the local maximum floating-point error. 
\section{Related Work}
\label{related-work}
In this section, we present research threads related to error detection in floating-point computations, including exception detection, condition numbers, and oracles computations of floating-point programs.

\subsection{Detecting floating-point exceptions}
Ariadne~\cite{barr2013automatic} modifies a numerical program to include explicit checks for each condition that can trigger an exception, performing symbolic execution using real arithmetic to find real-valued inputs that could induce an exception.  
NUMFUZZ~\cite{ma2022numfuzz} is a fuzzing tool that detects floating-point exceptions for numerical programs. It introduces a novel mutation strategy tailored for the floating-point format, designed to improve grey-box fuzzing by effectively generating valid floating-point test inputs. 

Our target numerical bugs are different from traditional exceptions. While exceptions like Overflow, Underflow, Inexact, Invalid, and Divide-By-Zero can typically be detected by compilers, certain bugs caused by large atomic condition numbers may only lead to inaccuracies, rather than triggering an exception. 
In cases where reliable oracles are unavailable, these inaccuracies can remain undetected.

\subsection{Conditioning}
To measure the inherent stability of a mathematical function, Higham et al.~\cite{higham2002accuracy} propose condition number, which could be derived using the Taylor series expansion. Since the condition number is hardly obtained for complex functions, study~\cite{fu2015automated} presents an automated method based on Monte Carlo Markov Chain techniques and mathematical optimization models to estimate condition numbers. Recently, Wang et al.~\cite{wang2022detecting} coarsely estimate condition numbers for real numerical programs and take them as fitness to guide the search.

\subsection{Computing the oracles of floating-point programs}
Oracles are indispensable for many subjects, such as detecting substantial errors in floating-point operations and searching for error-inducing inputs of programs. \textbf{1) With respect to detection:} FpDebug~\cite{benz2012dynamic} is designed based on MPFR~\cite{fousse2007mpfr} and Valgrind~\cite{nethercote2007valgrind}. This tool analyzes numerical programs by conducting all operations side-by-side in high-precision numerical representations. FPSanitizer~\cite{chowdhary2021parallel} is a novel prototype for parallel shadow execution to find errors comprehensively on the multicore machines. EFTSanitizer~\cite{chowdhary2022fast} further reduces the overheads by utilizing error-free transformation. \textbf{2) As for search:} Numerous tools~\cite{chiang2014efficient, zou2015genetic, yi2019efficient, guo2020efficient, wang2022detecting, zhang2023eiffel, zhang2024hierarchical} have to obtain oracles to help them guide search. They tend to perform floating-point computations in higher precision by using libraries such as \textit{mpmath}~\cite{mpmath} and MPFR~\cite{fousse2007mpfr}. 


To the best of our knowledge, ATOMU~\cite{zou2019detecting} and FPCC~\cite{yi2024fpcc} are the only tools that do not use the high-precision approach to guide search. ATOMU considers floating-point errors can be attributed to big condition numbers of atomic operations, such as catastrophic cancellation of subtraction and addition operations. They find inputs that result in huge condition numbers for each suspicious atomic operation by utilizing an evolutionary algorithm. However, some errors may be suppressed or masked during execution~\cite{bao2013fly}, thus it is not necessary and time-consuming to detect big condition numbers for all atomic operations. Although FPCC mitigates the issue of false positives, the computation of its chain condition number is time-consuming. In contrast, our approach injects perturbations at operations with big condition numbers and executes efficiently.

\section{Conclusion}
\label{conclusion}

Existing approaches typically employ high-precision programs to obtain oracles and calculate errors for detecting significant floating-point errors. However, the conventional solutions present two key limitations: difficulty of implementation and prolonged execution time. The two limitations substantially impair essential software engineering workflows, most notably in floating-point software verification and testing pipelines. Two recent tools, ATOMU and FPCC, have made progress in mitigating the limitations. However, ATOMU tends to produce false positives, whereas FPCC effectively eliminates them at the expense of significantly increased computational overhead. To address these issues, we propose a novel approach, termed PI-detector, which efficiently and effectively calculates floating-point errors. Our experimental results indicate that PI-detector can tackle all these aforementioned limitations. 

In future work, we intend to apply our approach to more professional and complex programs, such as numerical simulation applications, to assist scientists in assessing whether specific inputs could lead to significant errors. We are confident in our approach as it can handle multiple input data types while maintaining short execution times.

\bibliographystyle{IEEEtran}
\bibliography{ref}

\end{document}